\documentclass[aps,prl,reprint,groupedaddress]{revtex4-1}

\usepackage{amsmath}
\usepackage{braket}
\usepackage{graphicx}

\begin{document}

%\preprint{}

\title{Thermoelectric properties of two-dimensional Dirac materials}



\author{Eddwi H. Hasdeo}
\email[]{eddw001@lipi.go.id}
%\thanks{}
\affiliation{1Research Center for Physics, Indonesian Institute of Sciences, Kawasan Puspiptek Serpong 15314, Indonesia}

\author{Lukas P. A. Krisna}
\email[]{lukasprimahatva@live.com}
\affiliation{2Department of Physics, Institut Teknologi Bandung, Jl. Ganesha 10, Bandung 40132, Indonesia}

\author{Ahmad R. T. Nugraha}
\email[]{nugraha@flex.phys.tohoku.ac.jp}
\affiliation{Department of Physics, Tohoku University, 6-3 Aramaki-Aza-Aoba, Aoba-ku, Sendai 980-8578, Japan}

\date{\today}

\begin{abstract}
We performed Boltzmann transport calculation to obtain the Seebeck coefficient, electrical conductivity, electronic thermal conductivity, and thermoelectric figure of merit (ZT) for Dirac systems. We found an enhancement of ZT due to the gap opening. When the phonon thermal conductivity is small enough, the optimum ZT in gapped Dirac system can be larger than 1, which is preferable for thermoelectric applications.
\end{abstract}

%\keywords{}

\maketitle

\section{\label{sec:1-intro}Introduction}
Dissipative heat from engines is mostly wasted or circulated within the engine that induces overheating. To solve this problem, one can benefit from thermoelectric (TE) materials that are able to convert heat into electricity\cite{Chen2003,Zhang2015}. Thermoelectricity promises renewable energy harvesting as well as avoiding overheating in engines. The TE efficiency is described by the so-called “figure of merit”, $ZT=S^2\sigma T/\kappa$, where $\sigma$ is the electrical conductivity, S is the Seebeck coefficient, and $\kappa=\kappa_e+\kappa_{p}$  is the heat conductivity arising from electron contribution $\kappa_e$ and phonon contribution $\kappa_{ph}$\cite{Goldsmid2009}. This expression implies that good TE materials must conduct electricity well but poorly transport the heat. This is to keep the temperature gradients (on hot and cold sides) intact when carriers are transported. This requirement poses very challenging problem because carriers usually carry both charge and heat.   
So far, commercially available TE materials are bismuth telluride (Bi$_2$Te$_3$)\cite{Mamur2018}, lead telluride (PbTe)\cite{Dughaish2002}, and silicon germanium (SiGe)\cite{Joshi2008} with typical ZT of about 1 or corresponding efficiency 10\% in room temperature. Despite decades of research in thermoelectricity, ZT values remain stagnant for long years. Recent significant improvements of ZT  are inspired by theoretical works of Hicks and Dresselhaus that low dimensional materials, such as one-dimensional (1D) nanowires or two-dimensional (2D) nanosheets, improve the TE properties thanks to quantum confinement effects\cite{Hicks1993a,Hicks1993,Heremans2013}. This effect enables drastic change in the carriers’ density of states that enhances the Seebeck coefficient.   
Miniaturizations of materials, however, do not automatically enhance the TE properties. Some experiments reported that power factors $PF=\sigma S^2$ of nanomaterials are not enhanced compared with the bulk counterpart despite the 2D materials thicknesses and 1D materials diameters have reach nanoscale\cite{Boukai2008,Hochbaum2008,Kim2015}. We found out in our previous work that the confinement size L of materials (thickness in 2D or diameters in 1D) must be much smaller than thermal de Broglie wavelength $\Lambda$ to get the enhancement effect\cite{Hung2016,Hung2015}, where $\Lambda$ is inversely proportional to effective mass of electrons in solid. Thus, to obtain large ZT, we must use materials with small effective mass. 
Due to this confinement effect, atomic layered materials can be a strong candidate to enhance TE properties. Graphene, single-atom-thick layer of carbon, hosts electrons with effectively zero mass. The motion of electrons in graphene follows relativistic Dirac equations. However, graphene has no band gap thus its TE properties might be poor. There are Dirac materials with similar hexagonal structure like graphene but with gap, namely transition metal dichalcogenides (TMDs), which include, for example, MoS$_2$, MoSe$_2$, MoTe$_2$, WS$_2$, WSe$_2$, and WTe$_2$\cite{Geim2013,Novoselov2004,Wehling2014,Mak2010}. These materials are known to exhibit large TE power factor\cite{Hippalgaonkar2017}. Interestingly, their band gaps are electrically tunable when bias voltage is applied on the top and bottom of the materials\cite{Zhang2009}.
In this paper, we show calculation of Seebeck coefficient, electrical and thermal conductivities for both gapless and gapped Dirac systems.  By doing so, we can compare ZT of graphene and TMDs. We find that the presence of gap in TMDs may enhance ZT. Therefore, TMDs are more favorable than graphene to be used as TE materials.

\section{\label{sec:2-theo}Theoretical methods}
From the Boltzmann transport theory, by applying relaxation time approximation, the electrical conductivity $\sigma$, Seebeck coefficient $S$ and electron thermal conductivity $\kappa_e$ are given by:
	\begin{equation}
	\sigma=q^2\mathcal{L}_0,\quad S=\frac{1}{qT}\frac{\mathcal{L}_1}{\mathcal{L}_2},\quad \kappa_e=\frac{1}{T}\left(\mathcal{L}_2-\frac{\mathcal{L}_1^2}{\mathcal{L}_0}\right)
	\end{equation}
In Eq. (1), q is carrier’s charge and $L_i$ is the TE kernel\cite{Mahan1996}:
	\begin{equation}
	\mathcal{L}_i=\int_0^\infty\mathcal{T}(E)(E-\mu)^i\left(-\frac{\partial f}{\partial E}\right) dE
	\end{equation}
where $\mu$ is the chemical potential, reflecting the doping level, $f$ is the Fermi-Dirac distribution, and $\mathcal{T}(E)=v_x^2(E)\tau(E)\mathcal{D}(E)$ is the transport distribution function which depends on longitudinal velocity $v_x^2=v^2/2$, relaxation time $\tau$, and density of states $\mathcal{D}$.
The relaxation time corresponds to the travel of an electron before it scatters with impurities. This is described by Fermi’s golden rule, $\hbar/\tau=2\pi \sum_f|\braket{f|V|i}|^\delta(E_f-E_i)$, where $\braket{f|V|i}$is the impurity scattering matrix element from initial state i  to final state f. When impurity scattering potential is short range in real space, the matrix element can be regarded as a constant in k-space. One can assume $\tau=C\mathcal{D}(E)^(-1)$  with the density of states being $\mathcal{D}(E)=\sum_\textbf{k}\delta(E-E_k)$ and $C$ a constant.

\section{\label{sec:3-gapless}Thermoelectricity of gapless Dirac systems}

\begin{figure}[t]
 	\centering
 	\includegraphics[width=0.5\textwidth]{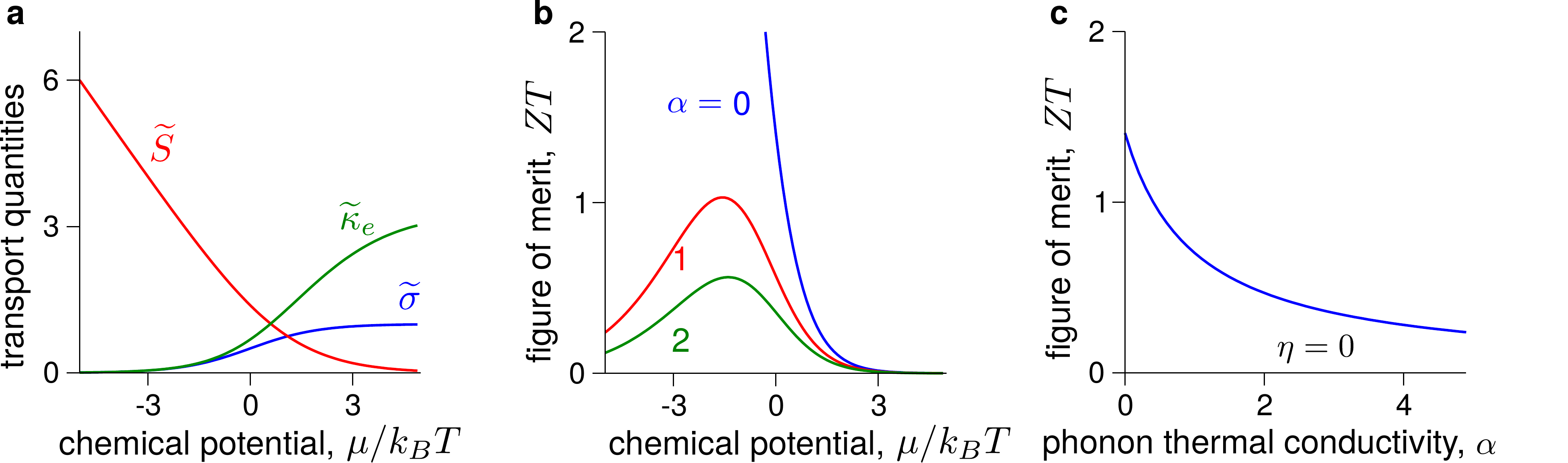}
 	\caption{(a) Electrical conductivity (blue) $\tilde{\sigma}=\sigma/\sigma_0$ with $\sigma_0=q^2 v^2 C/2$, Seebeck coefficient (red) $\tilde{S}=S/S_0$ with $S_0=k_B/q$ and electron thermal conductivity (green) $\tilde{\kappa}_e=\kappa_e/\kappa_0$ with $\kappa_0=v^2 Ck_B^2 T/2$ plotted as a function of chemical potential. (b) Figure of merit $ZT$ as a function of chemical potential for several values of phonon thermal conductivity $\alpha=\kappa_{ph}/\kappa_0$. (c) Figure of merit $ZT$ at zero doping $\eta=0$ as a function of phonon thermal conductivity $\alpha$.}
 	\label{fig:fig1}
 \end{figure}
 
We begin with defining a gapless single band Dirac fermion in two dimension with the energy dispersion $E=\hbar v|\textbf{k}|$, where $v=\frac{1}{\hbar}\lvert\frac{\partial E}{\partial \textbf{k}}\rvert$ is the electron’s group velocity and \textbf{k} is the wavevector in 2D Brilloun zone. The transport distribution function in gapless Dirac thus becomes constant: $\mathcal{T}=v^2 C/2$. The TE kernel for the gapless Dirac system can be expressed explicitly as:
	\begin{align}
	\mathcal{L}_i&=\frac{v^2C}{2}\int_{0}^{\infty}(E-\mu)^i\frac{1}{k_BT}\frac{e^{(E-\mu)/k_BT}}{(e^{(E-\mu)/k_BT}+1)^2}\,dE,\\
	&=\frac{v^2C}{2}(k_BT)^i\mathcal{F}_i(\eta),\quad\mathcal{F}_i(\eta)=\int_{\eta}^{\infty}dx(x)^i\frac{e^x}{(e^x+1)^2}
	\end{align}

where $\eta=\mu/k_B T$ is the reduced chemical potential and $k_B$ is the Boltzmann constant. Substituting Eq. (3) to (1), we can plot the $\sigma$, $S$, $\kappa_e$  as a function of doping $\mu$ in Figure~\ref{fig:fig1}a. We can see that $\sigma$ and $\kappa_e$  monotonically increase with the doping level. These two similar trends follow the Wiedermann-Franz law which reflects the fact that carriers carry both charge and heat. On the other hand, Seebeck coefficient decreases by increasing doping because $S$ is proportional to the rate change of $\sigma$ with respect to $E$.
Having those transport quantities, we can calculate the figure of merit $ZT=S^2 \sigma/(\kappa_e+\kappa_{ph})$ as shown in Figure~\ref{fig:fig1}b. Here, we do not calculate the phonon thermal conductivity but assume that $\kappa_{ph}=\alpha \kappa_0$ is a constant value over $\mu$ that can be obtained from experiment with $\kappa_0=v^2 Ck_B^2 T/2$. If $\kappa_{ph}$ is very small, $ZT$ can be theoretically very large as shown in blue line of Figure~\ref{fig:fig1}b. By increasing $\kappa_{ph}$, $ZT$ monotonically decreases as shown in Figure~\ref{fig:fig1}c. 
  
The measured values of $\kappa$ and  $\kappa_e$  are about 2500~5000 W/mK\cite{Ghosh2008,Cai2010,Chen2011} and 11 W/mK\cite{Yigen2013} respectively  which corresponds to a possible minimum $\alpha_{min}\approx155$ for $T=300 K$ at zero doping. This corresponds to $ZT = 0.006$ which is quite low compared with commercial TE materials. Gapless nature of graphene might hamper its TE properties despite the large electrical mobility. The enhancement in TE properties can be achieved in gapped Dirac materials as shown below.

\section{\label{sec:4-gapped}Thermoelectricity of gapped Dirac system}
Dirac system acquires gap due to inversion or time reversal symmetry breaking. Gapped Dirac system takes the effective energy band as $E(\textbf{k})=\sqrt{(\hbar v|\textbf{k}|)^2+\Delta^2}$, where $\Delta$ is the half of band gap. With this dispersion, the TE kernel becomes
	%\begin{widetext}
	\begin{align}	
	\mathcal{L}_i^\Delta&=\frac{Cv^2(k_BT)^i}{2}\left(\mathcal{F}_i(\eta-\tilde{\Delta})-\mathcal{G}_i(\tilde{\Delta},\eta)\right),\\
	\mathcal{G}_i(x,y)&=\int_{x-y}^{\infty}du\frac{x^2}{(u+y)^2}u^i\frac{e^u}{(e^u+1)^2},
	\end{align}
%\end{widetext}
where $\tilde{\Delta}=\Delta/k_BT$ is a reduced gap.
        
\begin{figure}[h]
  \centering \includegraphics[width=0.5\textwidth]{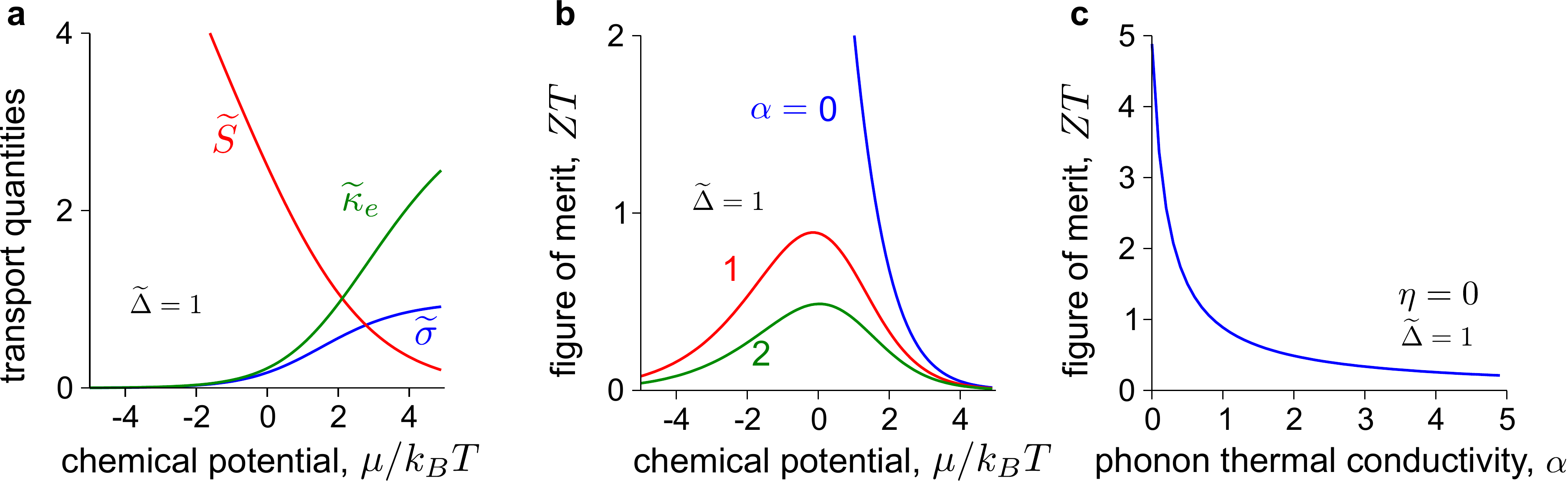}
  \caption{(a) Electrical conductivity (blue)
    $\tilde{\sigma}=\sigma/\sigma_0$ with $\sigma_0=q^2 v^2 C/2$,
    Seebeck coefficient (red) $\tilde{S}=S/S_0$ with $S_0=k_B/q$ and
    electron thermal conductivity (green)
    $\tilde{\kappa}_e=\kappa_e/\kappa_0$ of gapped Dirac systems
    plotted as a function of chemical potential with
    $\tilde{\Delta}=1$ and $\kappa_0=v2Ck_B^2T/2$. (b) Figure of merit
    $ZT$ of gapped Dirac system as a function of chemical potential
    for several values of phonon thermal conductivity
    $\alpha=\kappa_{ph}/\kappa_0$. (c) Figure of merit $ZT$ at zero
    doping $\eta=0$ as a function of phonon thermal conductivity
    $\alpha$.}
  \label{fig:fig2}
\end{figure}

Given the transport quantities in Figure~\ref{fig:fig2}a, we can plot $ZT$ for a
fixed gap $\Delta=k_BT$ for several values of phonon thermal
conductivity coefficient $\alpha$ in Figure~\ref{fig:fig2}b. The enhancement can be
seen that maximum $ZT$ reaches 5 at zero doping in Figure~\ref{fig:fig2}c (compare
with gapless case in Figure~\ref{fig:fig1}c maximum $ZT = 1.2$). This enhancement in
$ZT$ arises due to enhancement of Seebeck coefficient at low
doping. To estimate the $ZT$ value in MoS$_2$, thermal conductivity is
obtained to be 116.8 W/mK\cite{Jin2015} and 13.3 W/mK\cite{Bae2017}
with electronic contribution about two orders of magnitude
smaller. With gap $2\Delta = 1.79$ eV\cite{Kadantsev2012} ZT can reach
$0.15–1.20$ at $300$ K. Compare with previous $ZT$ calculation, our
obtained $ZT$ is comparable despite on the simplicity of the model we
used\cite{Jin2015}. Note that we do not consider the contribution from
hole (electrons in valence band) in the one-band model. For large-gap
semiconductors, this model works reasonably well. However, for
small-gap and gapless materials, hole contribution may cancel the
electron contribution hence reducing $ZT$. The multiband effect will
be discussed elsewhere.

\section{\label{sec:5-summary}Summary}
We have performed analytical calculation of Seebeck coefficient, electrical and thermal conductivities from electrons for gapless and gapped Dirac systems. We found that $ZT$ in gapped Dirac systems can be greater than 1, which is larger than conventional TE materials. Our calculation suggests that gapped Dirac system is prospective for TE devices and applications.

\end{document}